\documentclass[fleqn,usenatbib]{mnras}

\usepackage{newtxtext,newtxmath}

\usepackage[T1]{fontenc}
\usepackage{ae,aecompl}

\usepackage{graphicx}	
\usepackage{amsmath}	
\usepackage{amssymb}	
\usepackage{algorithm}
\usepackage{algorithmic}

\usepackage{xcolor}

\title[deep-REMAP]{deep-REMAP: Probabilistic Parameterization of Stellar Spectra Using Regularized Multi-Task Learning}

\author[S. Gilda]{
Sankalp Gilda$^{1}$\thanks{E-mail: sankalp.gilda@gmail.com. This paper has been accepted for publication in MNRAS.}
\\
$^{1}$DeepThought Solutions, FL, USA\\
}

\date{Accepted 08/10/2025. Received 30/09/2025; in original form 12/01/2024}

\pubyear{2025}

\begin{document}
\label{firstpage}
\pagerange{\pageref{firstpage}--\pageref{lastpage}}
\maketitle

\begin{abstract}
In the era of exploding survey volumes, traditional methods of spectroscopic analysis are being pushed to their limits. In response, we develop deep-REMAP, a novel deep learning framework that utilizes a regularized, multi-task approach to predict stellar atmospheric parameters from observed spectra. We train a deep convolutional neural network on the PHOENIX synthetic spectral library \citep{phoenix_grid} and use transfer learning to fine-tune the model on a small subset of observed FGK dwarf spectra from the MARVELS survey \citep{marvels_ge_intro}. We then apply the model to 732 uncharacterized FGK giant candidates from the same survey. When validated on 30 MARVELS calibration stars, deep-REMAP accurately recovers the effective temperature ($T_{\rm{eff}}$), surface gravity ($\log \rm{g}$), and metallicity ([Fe/H]), achieving a precision of, for instance, approximately 75 K in $T_{\rm{eff}}$. By combining an asymmetric loss function with an embedding loss, our regression-as-classification framework is interpretable, robust to parameter imbalances, and capable of capturing non-Gaussian uncertainties. While developed for MARVELS, the deep-REMAP framework is extensible to other surveys and synthetic libraries, demonstrating a powerful and automated pathway for stellar characterization.

\end{abstract}




\section{Introduction}\label{sec:introduction}

In the last few decades, computers have revolutionized astronomy. The emergence of novel and complex instruments, advances in telescope designs allowing for ever larger apertures, and increased private and public funding have enabled us to not only exponentially expand survey volumes, but also to observe objects with unprecedented resolution. 
Just within the the Sloan Digital Sky Survey (SDSS, \citealt{york2000sloan, eisenstein2011sdss}), various large-scale sky survey programs such as the Sloan Extension for Galactic Understanding and Exploration (SEGUE, \citealt{segue}), RAdial Velocity Experiment (RAVE, \citealt{rave}), Baryon Oscillation Spectroscopic Survey (BOSS, \citealt{boss}), and the Large Sky Area Multi-Object Fiber Spectroscopic Telescope Telescope (LAMOST, \citealt{lamost}) have collected spectra for hundreds of thousands to millions of stars. Massive data releases like those from the Gaia-ESO Survey \citep{gaiaeso}, the Dark Energy Survey Instrument (DESI, \citealt{desi}), and the Large Synoptic Sky Survey (LSST, \citealt{lsst}) will not only add to the corpus of data, but will demand significant computational resources for subsequent analysis. Modern methods in machine learning and statistics are playing an increasingly important role in virtually all fields of astronomy---deep learning (DL) frameworks have been developed and utilized in star-galaxy classification \citep{star_galaxy_classification}, photometric redshift prediction \citep{dnn_phot_redshift}, identification of galaxy-galaxy strong lensing \citep{strong_lensing}, and analysis of survey data products \citep{ztf}, to name a few applications.

Specifically with respect to spectral analysis, this new wave of automation has offered many alternatives to help overcome limitations of the conventional methods. Accurate stellar characterization is essential not only for understanding stellar evolution but also for applications such as exoplanet detection via radial velocity methods \citep{ge_etal_exo_detection}, where stellar parameters directly impact derived planetary properties.

One such widely used method for determination of stellar atmospheric parameters from absorption spectra (henceforth `stellar parameterization') is the `Equivalent Widths (EW) Method', detailed in \citet{equivalent_width}. Although a powerful technique for high resolution data \citep{valenti2005spectroscopic}, it fails to parameterize low to moderate resolution spectra due to the blending of spectral lines. The commonly used technique in such cases is spectral synthesis, which involves the determination of stellar parameters through a comparison of an observed spectrum with an extensive library of synthetic ones. This has been widely used in data analysis pipelines for several surveys, such as SEGUE \citep{spec_syn_segue_1,spec_syn_segue_2}, RAVE \citep{rave}, LAMOST \citep{lamost} and AMBRE \citep{spec_syn_ambre_1, spec_syn_ambre_2}. Some of the drawbacks of this method \citep{ghezzi2014accurate} include a dependency on the completeness and accuracy of atomic line databases, and the need to accurately determine broadening parameters (instrument profile, macro turbulence, and rotational velocities). To combat this issue, \citet{ghezzi2014accurate} developed and utilized spectral indices---specific spectral regions combining multiple absorption lines into broad, blended features. However, this method has its own set of limitations, primarily that for a given survey, it requires the user to calculate and input a list of spectral indices and their respective equivalent widths. These methods, in addition, are also data-inefficient, in that they only consider certain parts of a given spectrum while assigning little importance to the rest.

Successful and efficient interpretation of large-scale, high-dimensional spectral data, in the presence of the above mentioned limitations of the conventional methods, has forced us to 
embrace advances in informatics. The past decade and a half alone alone has seen the development of several machine learning (ML) based algorithms for stellar parameterization---from simple ones that use multi layer perceptrons (MLPs; simple sequences of fully connected layers, \citep{mlp_1,mlp_2,mlp_3}) to more sophisticated ones based on convolutional neural networks (CNNs), such as ``The Cannon'' \citep{cannon}, ``The Cannon 2'' \citep{cannon2}, ``The Payne'' \citep{payne}, ``StarNet'' \citep{starnet}, \citet{dnn_example_stellar_param}'s work on stellar parameter prediction using Principal Component Analysis, and \citet{sivarani}'s work on predicting stellar spectral type by leveraging semi-supervised learning, and \citet{gilda2021mirkwood} who developed an ML framework for SED modeling. While promising steps in the right direction, these works either use simple networks incapable of capturing the complicated relationships between spectra and stellar parameters, require large amount of already labeled data from a given survey to parameterize newly observed stars, and/or are restricted to the range of stellar parameters available for the training data (i.e., do not generalize to out of sample distributions). In this work, we develop $\rm{deep-REMAP}$---a novel, interpretable, state-of-the-art neural network capable of overcoming these limitations. We maximize its performance by leveraging the most advanced training practices, routines, and loss functions in the deep learning (DL) literature. We apply our network to one-dimensional (1D) spectra from the MARVELS survey \citep{marvels_ge_intro}, validate its skill on stars with known atmospheric parameters, and for the first time predict these values (T$_{\rm{eff}}$, $\rm{log\;g}$, $\rm{[Fe/H]}$) for 732 FGK giant star candidates.

Recent advances have begun addressing the synthetic-observational mismatch more systematically. \citet{obriain2021} introduced Cycle-StarNet, using unsupervised domain adaptation to bridge the gap between synthetic and observed spectra. For low-resolution spectra, \citet{li2024} developed AspGap, successfully applying transfer learning from APOGEE to Gaia XP spectra. \citet{leung2023} demonstrated transfer learning's efficacy for spectroscopic age estimation. However, moderate-resolution spectra like MARVELS (R$\sim$11,000) present unique challenges -- line blending precludes traditional EW methods while the resolution remains insufficient for standard high-resolution ML approaches. This necessitates the careful combination of transfer learning and probabilistic frameworks we present in this work.

The rest of this paper is organized as follows: In \S\ref{sec:dcnn} we briefly describe convolutional neural networks with a specific focus on the architecture of $\rm{deep-REMAP}$. In \S\ref{sec:data} we characterize both MARVELS and synthetic PHOENIX spectral data, and delineate the pre-processing steps undertaken before both sets of spectra are input into the neural network. In \S\ref{sec:transfer learning} through \S\ref{sec:swa}, we outline the various state-of-the-art modules and training practices used in this paper to improve our network's performance. In \S\ref{sec:implementation details} we sketch out in detail our training routine, loss functions, and model hyperparameters. In \S\ref{sec:results} we showcase $\rm{deep-REMAP}$'s performance on predicting the three stellar parameters, both for the synthetic and real MARVELS spectra. We also breakdown the model skill as a function of the various methods described in \S\ref{sec:methodology}, thus highlighting their importance and justifying their inclusion in our model. Finally, in \S\ref{sec:conclusions} we summarize our work and suggest possible routes of inquiry for future research endeavors in this space. In brief, our key contributions are as follows.
\begin{enumerate}
    \item We demonstrate the necessity and successful implementation of transfer learning to better map synthetic libraries to real observations.
    \item We utilize and demonstrate the effectiveness of many state-of-the-art methods increasingly common in several applications of deep learning, but missing in the astronomical literature.
    \item We show that by converting a regression problem to a classification one, switching the softmax loss function to one better able of handing imbalanced classes, and regularizing it with an embedding loss, our network is not only able to work with random probability distributions, but is also interpretable by being able to retrieve nearest neighbors for any given query spectrum.
    \item We predict, for the first time, the stellar parameters (and related uncertainties) for 732 suspected MARVELS red giant stars.
\end{enumerate}

\section{Convolutional Neural Networks}\label{sec:dcnn}
Convolutional Neural Networks (CNNs, e.g. \citealt{lecun1998, imagenet2012, inception2016}) are a class of artificial neural networks used in image recognition and processing that are specialized for application on pixel and voxel-based data. They consist of an input layer, an output layer, and one or more hidden layers (a CNN with more than one hidden layer is referred to as a Deep CNN, or a DCNN). Over many cycles (or \emph{epochs}) the network learns the convolutional filters, weights (or \emph{kernels}), and biases necessary to extract meaningful patterns from the input image. For an input image, a DCNN assigns importance to various aspects/objects in it, and tries to map them to the output label of choice in a way that minimizes the user-defined \emph{loss-function}. Compared to other image processing algorithms, DCNNs require very little pre-processing of the input images (such as creation of hand-crafted features); with enough training, the filters of a DCNN have the ability to extract relevant features for classification or regression tasks. Through the application (\emph{activations}) of its filters, a DCNN is thus able to successfully capture the spatial dependencies in an image. It has fewer parameters and more sophistication than a multi-layer perceptron, and is a far superior choice for handling images, especially at scale. (See also \citealt{dnn_intro1}, \citealt{bengio2017deep}, and references therein for a comprehensive introduction to neural networks).

\begin{figure*}
\centering
\includegraphics[width=1\linewidth]{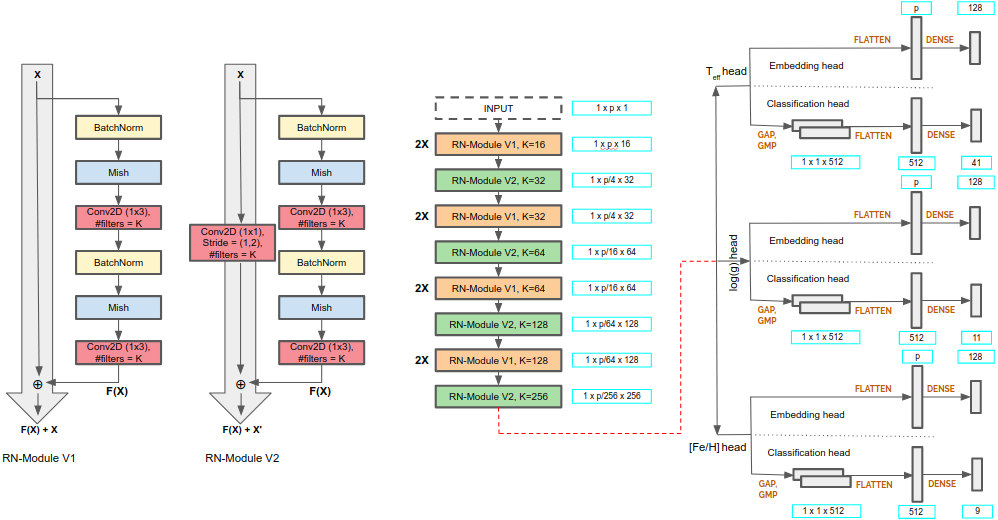}
\caption{Schematic of the $\rm{deep-REMAP}$ neural network architecture. \textbf{Left:} The structure of the two types of residual modules used, RN-Module V1 (identity) and V2 (projection for downsampling). \textbf{Middle:} The network backbone, consisting of a sequence of residual modules that progressively increase the feature depth (K) while reducing the spectral dimension (p). \textbf{Right:} The multi-task head structure. The shared feature backbone splits into three parallel heads, one for each stellar parameter ($T_{\rm{eff}}$, $\log \rm{g}$, [Fe/H]). Each head further splits into an \emph{embedding head} (trained with triplet loss) to structure the latent space, and a \emph{classification head} (trained with focal loss) to produce the final probabilistic prediction.}
\label{fig:deep_remap}
\end{figure*}

The $\rm{deep-REMAP}$ architecture is illustrated in Figure \ref{fig:deep_remap}. It consists of several \emph{residual modules}, batch-normalization layers and dropout layers for regularization \citep{batchnorm, dropout}, and parallel mean and max pooling layers for feature extraction and reduction of overfitting \citep{pooling}. Residual modules were introduced in \citet{resnet} as part of the widely used $\rm{ResNet}$ architecture. Depicted in Figure \ref{fig:deep_remap} (left panel), the residual module comes in two flavors: the \emph{identity residual module} and the \emph{projection residual module}. The former is a block of two convolutional layers with the same number of filters and a small filter size, where the output of the second layer is added with the input to the first convolutional layer. The latter uses strides to reduce the size of the input image, and adds more filters. We also use three sets of \emph{heads} for extracting features specific to each of the three parameters, and making predictions using them---all while taking the parameters' mutual interdependence into account. Each of these heads in turn consist of two sub-heads for two different types of losses---a \emph{classification loss} and an \emph{embedding loss} (see \S\ref{sec:methodology} for details).

In addition, we also identify a few choices that are critical in obtaining the results reported in \S\ref{sec:results}:
\begin{enumerate}
    \item We use the state-of-the-art Mish activation function \citep{mish_act}. We find that this significantly speeds up training compared to both ReLU \citep{relu1,relu2,relu3} and Leaky ReLU \citep{leakyrelu}.
    \item We use both global average pooling and global max pooling. We find that using both, we are able to capture more informational deep features than with either one of the pooling schemes alone.
\end{enumerate}

For each of the three stellar parameters, we use the focal-loss \citep{focal_loss} as the classification loss, and the triplet-loss \citep{triplet_introduction,triplet_face_recognition,triplet_deep_metric_learning,triplet_person_reidentification} as the embedding loss. We iteratively update the model weights using the `Adam' optimizer \citep{adam_opt}. We bin the real-valued variables T$_{\rm{eff}}$, $\log \rm{g}$, and [Fe/H] using the median values of the distributions of their respective uncertainties, and use \emph{one-hot encoding} (aka \emph{one-of-K encoding}) to convert those real-valued bins into binary values\footnote{For instance, if one were aiming to classify the color of a flower as either white, red or orange, one version of \emph{one-hot encoding} these would be: white$\rightarrow$[1,0,0], red$\rightarrow$[0,1,0], and orange$\rightarrow$[0,0,1]}. The output predictions are converted back to numerical bins. Details about one-hot encoding and decoding, loss function creation, and the training scheme and are discussed in \S\ref{sec:methodology}.


\section{Data and Pre-Processing}\label{sec:data}

\subsection{MARVELS Spectra}\label{sec:marvels spectra}
In this work we use spectra from the Sloan Digital Sky Survey III (SDSS-III, \citealt{eisenstein2011sdss}) Multi-object APO Radial Velocity Exoplanet Large-area Survey (MARVELS, \citealt{marvels_ge_intro}) taken with the SDSS 2.5-m telescope at the Apache Point Observatory \citep{marvels_telescope}. Between October 2008 and July 2009, using a fibre-fed dispersed fixed delay interferometer (DFDI) and a medium resolution (R$\sim$11,000, \citealt{marvels_ge_2}) spectrograph, MARVELS observed $\sim$5,500 stars with a goal of characterizing short-to-intermediate period giant planets in a large and homogeneous sample of FGK dwarfs. \citet{grieves2017exploring} compared the MARVELS radial velocity results from the University of Florida Two Dimensional (UF2D, \citealt{thomas_dissertation}) data processing pipeline to results from the University of Florida One Dimensional (UF1D, \citealt{marvels_pipeline_thomas}) pipeline, while \citet{sdss_dr11_and_12} presented an overview of previous MARVELS data reductions. \citet{Paegert_2015} describes the MARVELS target selection process, along with the ineffective giant star removal method which resulted in $\sim$24\% giant contamination rate.

As mentioned in \S\ref{sec:introduction}, \citet{ghezzi2014accurate} developed the spectral indices method as an alternative approach to the spectral synthesis technique to obtain accurate atmospheric parameters for low to moderate resolution spectra. They tested the method specifically for MARVELS with a validation sample of 30 MARVELS stars that had high resolution spectra obtained from other instruments. 
Each MARVELS star has two sets of spectra due to the interferometer which creates two ``beams''. Each set of spectra are analyzed separately and two sets of parameters for each star are combined using a simple average and uncertainties are obtained through an error propagation. \citet{grieves2018chemo} expanded the application of this method further and derive stellar parameters for all $\sim$5,500 observed stars. They applied certain cuts based on number of observations per star ($\geq$ 10), quality of observations (poor or missing observations caused by dead or mis-plugged fibers, spectra suggesting the star is a spectroscopic binary, or unreasonable photon errors), and uniqueness of stars (they identify 7 duplicates). This culling leaves 3,075 unique stars, which are then further divided into dwarfs+subgiants and giants. This is because the spectral indices pipeline was designed specifically for a certain range of stellar parameters---4,700 K $\leq$ T$_{\rm{eff}}$  $\leq$ 6,000 K, 3.5 $\leq\log\;\rm{g} \leq$ 4.7, and -0.9 $\leq$ [Fe/H] $\leq$ 0.5. The resulting 2,540 dwarfs+subgiants were subsequently winnowed down to 2,343 dwarfs based on the definition from \citet{ciardi_kepler} (see \S 3 in \citealt{grieves2018chemo} for more details).

We use the stellar parameters (and associated uncertainties) of 2,313 of these stars (Nolan Grieves, private communication) to fine-tune our network, validate its performance on the remaining 30 calibration stars, and use it to predict the parameters for the residual 732 giant and sub-giant candidate stars.

\subsection{PHOENIX Spectra}\label{sec:synthetic spectra}
The PHOENIX grid of synthetic spectra \citep{phoenix_grid} was created for high resolution (R $\geq$ 100,000) stellar spectra spanning ultra-violet to infrared wavelengths (50--5,000 nm). It spans a large parameter space (2,300 $\leq$ T$_{\rm{eff}}$ $\leq$ 12,000 K, 0 $\leq\log\;\rm{g}\leq$ 6, -4 $\leq$ [Fe/H] $\leq$ 1, -0.2 $\leq$ [$\alpha$/Fe] $\leq$ 1.2). 
In this work, we consider only the subset of PHOENIX grid relevant to the MARVELS parameter space. Since MARVELS stars are expected to be F,G and K type stars, we limit the PHOENIX stellar parameter to: 4,850 $\leq\lambda\leq$ 5,750 \AA, 3,000 $\leq$ T$_{\rm{eff}}$ $\leq$ 7,000 K, 1.0 $\leq \log\;\rm{g} \leq$ 5.5, -2 $\leq$ [Fe/H] $\leq$ 1. Below, we describe how the synthetic PHOENIX spectra are brought in line with MARVELS spectra to reduce the discrepancies between them, thus allowing for successful transfer learning.

\subsection{Pre-processing and Data Augmentation}\label{sec:preprocessing}
Synthetic spectra are typically created to either be commensurate with observed spectra from a particular survey, or at least to reproduce their major absorption features. For instance, the INTRIGOSS \citep{intrigoss} synthetic grid was tuned by direct comparison to Gaia-ESO spectra of FGK stars. Similarly, \citet{phoenix_grid} used PHOENIX to analyse MUSE integral field spectra of stars in the metal-poor globular cluster NGC 6397. Thus to be able to use such a synthetic grid for analysis of real observed spectra for which it was not designed, we have to take special care to traverse the \emph{synthetic gap}. This refers to the differences in the shapes, depths, and number of absorption lines between the two sets, arising due to both instrumental and observational effects. To this end, we make the following changes to the the subset of PHOENIX spectra selected in \S\ref{sec:synthetic spectra}, in the described order:
\begin{enumerate}
    \item Down-convolution: We convolve PHOENIX spectra with three Gaussian kernels to reduce their resolution from 100,000 to 10,000, 11,000 and 12,000; this is to account for the resolution-variation across CCD elements. 
    \item Resampling: We resample PHOENIX spectra onto the wavelength grid of MARVELS via cubic spline interpolation; this reduces the wavelength spacing from 0.01 to 0.1 \AA.
    \item Noise Addition: We add Gaussian noise to all PHOENIX spectra with the constraint $0.5\% \leq \sigma \leq 2\%$, corresponding to the Signal-to-Noise (SNR) distribution of the majority of MARVELS spectra ($50 \leq$ SNR $\leq 200$). At such high values of SNR, Gaussian noise is a very good substitute for Poisson noise. We uniformly choose five different values of $\sigma$ in the above range, and sample thrice for each of the five values.
    \item Continuum Normalization: We remove the continuum from both PHOENIX and MARVELS data via the same method (described below) for consistency.
\end{enumerate}

All the above steps constitute \emph{pre-processing}---making synthetic data ready for use with $\rm{deep-REMAP}$---while steps (i) and (iii) also constitute \emph{data augmentation}. The latter is a very common method used in ML and DL applications, where we create new samples to both increase the size of the training set (deep neural networks are notoriously data hungry), and to make the model robust to variations expected in real data. Since we know that MARVELS spectra have varying resolutions with wavelength, and varying noise levels due to different number of spectra stacked to produce the final spectrum for each star, we introduce this information in the `vanilla' PHOENIX spectra.
\\

\noindent \textbf{Continuum Normalization}: Continuum estimation and removal is a key component in spectroscopic analysis. While several methods exist in literature---polynomial fitting to either the whole or part of the spectrum \citep{cannon2, cont_norm_poly}, sigma clipping \citep{starnet}, smoothing with a Gaussian kernel \citep{cont_norm_gaussian}---in this work we develop a novel routine (described in Algorithm \ref{algo:cont_norm}). Extensive testing on synthetic spectra spanning a large parameter space, and injected with various levels of noise, demonstrated the effectiveness of our algorithm. In future, we plan on comparing the performance and efficiency of our continuum normalization scheme with more advanced algorithms in literature, such as those based on the wavelet transform \citep{cont_norm_wavelet}, the framelet transform \citep{cont_norm_framelet}, and a combination of the wavelet transform with Kalman filtering \citep{cont_norm_wavelet_kalman} for simultaneous signal denoising and continuum removal.  

Like with any real spectra, the 3,075 unique MARVELS spectra need careful consideration before we can accurately estimate their continuum levels. Firstly, we identify and remove traces of \emph{fake features} from the combined 2D spectra before collapsing them to 1D. These include cosmic rays and interferometric signatures: $\sim$20\% MARVELS spectra contain bright features that can span up to six pixels ($\approx 1$ \AA) in width \citep{grieves2017exploring, grieves2018chemo}. These features remain stationary (in wavelength space) throughout the time-span of the survey, which simplifies their removal. Identification and removal of these dummy/fake fluxes is critical as saturated flux values can negatively impact continuum finding routines. After determining the wavelength solution and de-Dopplerizing the spectra to bring them to a rest frame, we then apply our novel continuum finding routine described in Algorithm \ref{algo:cont_norm}. An important point to note is the impact that low signal-to-noise ratios have on continuum fitting---noisy flux values in a spectrum can easily be confused as true and result in a higher estimation of the continuum compared to its true value. While MARVELS spectra have varying SNRs---both due to the different number of observations stacked to create the final spectrum for a given star, and the fact that the flux for a given star varies widely for different observations due to sky conditions and fibre degradation---our experiments show that for their median value of $\sim$200, this is only a second-order effect. 
We therefore run Algorithm \ref{algo:cont_norm} by assuming an SNR of 200 for all the MARVELS spectra. At the same time we incorporate the variation in SNRs across spectra of MARVELS stars by adding Gaussian noise to PHOENIX spectra before normalization (step (iii) above). Finally, we note that our method is subject to the same limitation as other methods of continuum normalization, 
in that as we move to lower temperatures, the estimated continuum moves below the true continuum (see Figure 2 in \citealt{starnet}). These considerations make it critical that the same continuum fitting routine is applied to both synthetic and real spectra, to ensure that both are subject to the same biases.

With the above steps completed, we end up with a total of 110,000 PHOENIX spectra, and 3,075 MARVELS spectra.

\begin{algorithm}
\caption{Continuum Normalization}
\label{algo:cont_norm}
\begin{algorithmic}[1]
\renewcommand{\algorithmicrequire}{\textbf{Input:}}
\renewcommand{\algorithmicensure}{\textbf{Output:}}
\REQUIRE a vector of spectrum flux measurements of length \emph{N}, $\textbf{Z} = \{z_n\}^{N}_{n=1}$.
\ENSURE the normalized spectrum flux vector of length \emph{N}, $\textbf{X} = \{x_n\}^{N}_{n=1}$.
\STATE Set $a$ as the maximum quantity of flux values in each bin that are allowed to be larger than the continuum.
\STATE Define the initial continuum estimate, $\textbf{C}$, as a vector of length $N$ with all elements set to 10 + max($\textbf{Z}$).
\FOR{bin count $v$ in range [2, 11]} 
\STATE Partition the index vector, $[1, N]$, into $v$ uniform bins. In each partition, determine the center element and place into vector $\textbf{H}$, $\textbf{H} = [N/v/2, \{[N/v/2, N]\}^{N}_{n=1+N/v}]$
\WHILE{the quantity of $\TRUE$ values in $\textbf{C} > \textbf{Z}$ is greater than $a$, }
\FOR{bin index $h$ in a random permuted vector of $\textbf{H}$} 
\STATE Create a subset of vector $\textbf{C}$ by duplicating indices $\textbf{H}$ of $\textbf{C}$ into vector $\textbf{E}$.
\IF{$\textbf{C}(h) > \textbf{Z}(h)$}
\STATE Lower the continuum estimate in bin $h$ by multiplying $\textbf{E}(h)$ by  0.999.
\IF{$v$ < 5}
\STATE Calculate the continuum estimate, $\textbf{C}$, by fitting a $v - 1$ degree polynomial to $(\textbf{H}, \textbf{E})$ and evaluate at range $[1, N]$.
\STATE To avoid undesired inflections at the continuum edges we perform linear interpolations to force flat edges in the endpoint bins. At the left edge we find the slope between the points $\textbf{C}(\textbf{H}(1))$ and $\textbf{C}(\textbf{H}(1) + 1)$, then extrapolate to replace the index range $[1, \textbf{H}(1) - 1]$. The right edge is altered in the same manner at the index range $[\textbf{H}(v) + 1, N]$.
\ELSE{}
\STATE Calculate the continuum estimate, $\textbf{C}$, by linear interpolation of $(\textbf{H}, \textbf{E})$ and evaluate at range $[1, N]$.
\ENDIF
\ENDIF
\ENDFOR
\ENDWHILE
\ENDFOR
\STATE Calculate the continuum estimate, $\textbf{C}$, by fitting a $6$ degree polynomial to $(\textbf{H}, \textbf{E})$ and evaluate at range $[1, N]$.
\STATE To avoid undesired inflections at the continuum edges we perform linear interpolations to force flat edges in the endpoint bins. At the left edge we find the slope between the points $\textbf{C}(\textbf{H}(1))$ and $\textbf{C}(\textbf{H}(1) + 1)$, then extrapolate to replace the index range $[1, \textbf{H}(1) - 1]$. The right edge is altered in the same manner at the index range $[\textbf{H}(v) + 1, N]$.
\STATE To smooth out the linear interpolations performed at the edges the continuum estimate is convolved with a gaussian kernel with sigma of 150.
\end{algorithmic} 
\end{algorithm}

\section{Theoretical Background}\label{sec:methodology}
We briefly describe here several advanced deep learning tools that are used in $\rm{deep-REMAP}$ to achieve the results in \S\ref{sec:results}.

\subsection{Transfer Learning}\label{sec:transfer learning}
In recent years, deep neural networks have become increasingly good at mapping from inputs to outputs, whether they are images, sentences, or time series. However, this paradigm of \emph{supervised learning}, by virtue of its dependence on large amounts of labeled inputs, is not only data hungry, it is also limiting in the ability to generalize to conditions or data samples that are different from the ones encountered during training. This type of learning is by design isolated and occurs purely based on specific tasks and datasets, and training different, isolated models on each. No knowledge is retained which can be transferred from one model to another.

\emph{Transfer learning} and related domain adaptation techniques \citep{gilda_unsupervised_domain_adaptation} allow us to deal with these scenarios by leveraging the already existing labeled data of some related task or domain. We store the knowledge gained from solving the \emph{source task} using the \emph{source data}, and apply it to solve the \emph{target task} of interest using \emph{target data}. This allows us to tackle the dual problems of sparsity and incompleteness in the \emph{target data}. In this work, we use transfer learning to probabilistically predict stellar atmospheric parameters for suspected giant stars in the MARVELS survey.

While several details of how CNNs work so well are still a matter of active research \citep{zeiler2014visualizing,olah2017feature,bau2017network,kim2018interpretability}, one thing that we are sure of is that lower layers capture low-level image features such as edges, while higher layers focus more on complex details, such as specific compositional features \citep{imagenet2012}. In other words, lower layers capture general representations, while higher layers capture dataset specific information. It then follows that for training a DCNN on different (but similar in at least some way) dataset than the one on which it was trained,\footnote{For instance, a neural net for classifying photographs of people into males and females, can be re-tooled to classify photographs of people into different age groups.} we can re-use the already extracted features. In practice, we update the weights and biases of the `pre-trained' model with a small learning rate in order to ensure that we do not unlearn the previously acquired knowledge. This simple approach of \emph{fine-tuning} has been shown to achieve astounding results on an array of vision tasks \citep{yosinski2014transferable,razavian2014cnn,girshick2014rich,long2015fully}.

In this work, we train $\rm{deep-REMAP}$ on a large corpus of augmented synthetic PHOENIX absorption spectra that have been pre-processed to match MARVELS characteristics (see \S\ref{sec:preprocessing} for details). Once the network achieves convergence (i.e., error on the validation set no longer decreases), we retrain it with a very small layer-dependent learning rate\footnote{Shallow layers are fine-tuned with the lowest learning rates since they capture the `big details' which are likely to be similar between both PHOENIX and MARVELS spectra; the learning rate is linearly increased with progressively higher layers. See \S\ref{sec:implementation details} for specific values.} on labeled MARVELS spectra (F,G and K-dwarfs for which we know the atmospheric parameters using spectral indices from \citealt{grieves2018chemo}) until convergence. Finally, this re-trained/fin-tuned network is used to make predictions on suspected 732 giants in the MARVELS survey.

\subsection{Multi-Task Learning}\label{sec:mtl}
In traditional deep learning, we typically create and train a neural network to optimize performance on one particular task---differentiating cats from dogs, determining the presence or absence of a tumor in an MRI image, predicting the type of solar flares based on spatio-temporal data of the Sun. While it can achieve impressive results on a host of tasks, this paradigm of learning has two disadvantages: (i) to optimize for multiple tasks, we are forced to chain multiple models, each optimized for a single goal; this increase in the number of models leads to a large number of parameters needing to be trained, which increases the chance of over-fitting; and (ii) by focusing on a single task at a time, we ignore shared information that might be prevalent across different tasks, and reduce the generalization ability of our models. An alternative to such \emph{single-task learning} approach is to share representations across inter-related tasks, and thus to enable our model to generalize better on on all tasks \citep{ multitask_wiki_2, caruana1997multitask,multitask_wiki_1}. This is the sub-field of \emph{Multi-Task learning} (MTL), which lets us exploit and exploit the commonalities and differences across related tasks.

\citet{caruana1997multitask} characterized MTL as, ``... an approach to inductive transfer that improves generalization by using the domain information contained in the training signals of related tasks as an inductive bias. It does this by learning tasks in parallel while using a shared representation; what is learned for each task can help other tasks be learned better.'' Inductive bias forces a model to preferentially select some hypotheses over others. For example, a common form of inductive bias is \emph{l1} regularization, which leads to a preference for sparse solutions. In the case of MTL, the inductive bias is provided by the auxiliary tasks, which cause the model to prefer hypotheses that explain more than one task; this generally leads to solutions that generalize better \citep{ruder2017overview}.

While MTL has showed remarkable success across a number of fields---natural language processing \citep{multitask_nlp}, computer vision \citep{multitask_fasterrcnn} and cancer discovery \citep{multitask_cancer}---to the best of our knowledge, this is the first work utilizing MTL in the analysis of stellar spectra. In this paper, we interpret each `task' as the prediction of one of the three stellar parameters, and train our model to simultaneously classify 1D absorption spectra into appropriate bins of effective temperature (T$_{\rm{eff}}$), surface gravity ($\rm{log\;g}$), and metallicity ([Fe/H]). We provide the implementation details in \S \ref{sec:implementation details}.

\subsection{Asymmetric Loss Function}\label{sec:assym loss func}
In this work, we have designed $\rm{deep-REMAP}$ as a classifier (for predicting discrete labels) as opposed to a more intuitive regressor (for predicting continuous, real-valued labels) for two reasons: (i) it makes it easy to generalize the network to work with non-Gaussian distributions of labels (i.e., unlike in \citealt{starnet} we are no longer restricted to predicting just the mean and standard deviation of the three labels T$_{\rm{eff}}$, [Fe/H], and $\log\;\rm{g}$); and (ii) it makes it easy to apply an embedding loss such as the triplet loss, and thus makes the model interpretable by enabling the user to query the nearest neighbors in the embedding space. This allows for qualitative assessment of which known spectra the model considers most similar to the input. Once we have discretized the three stellar parameters for use by the model, we are still left with two types of asymmetries:
\begin{enumerate}
    \item Arising from imbalanced training data. This refers to the fact that class distributions of the three atmospheric parameters for MARVELS stars are not uniform, which can bias the model towards the label of the majority class. For example, in the absence of any corrective mechanisms, a neural network fine-tuned on such a dataset would be more likely to deduce that the temperature of a previously unseen spectrum is 5800 K than 4200 K or 6900 K. 
    \item Arising from finite, Gaussian distribution of the atmospheric parameters about their mean. By default, classification losses assume a Dirac-delta like distribution of output labels. In other words, when a network with such a loss function sees a stellar spectrum in the training set with temperature 5,000 K, it will try to predict this by assigning a weight of 1.0 to the temperature bin corresponding to 5,000 K, and penalize all other temperature bins equally by assigning them the same weight of 0.0. This is in direct contrast to the Gaussian distribution of properties observed in nature---we know that for the above star, a temperature of 4,850 K is much more likely than temperature of 6,850 K, and hence the bins corresponding to these two temperatures should be penalized differently.
\end{enumerate}
To handle (i) above, we use \emph{focal loss} \citep{focal_loss} instead of cross-entropy loss. Focal loss intrinsically handles class imbalance by forcing the model to focus more on incorrectly predicted samples and less on correctly predicted ones. To handle (ii), we introduce a cost matrix that asymmetrically penalizes incorrect labels by assigning Gaussian weights, as expounded upon in \S\ref{sec:implementation details}.

\subsection{Triplet Loss}\label{sec:triplet loss}
The most commonly used loss functions, cross-entropy loss for classification and mean squared error loss for regression, aim to predict a discrete or continuous-valued label given an input. While modern convolutional architectures (such as $\rm{ResNet}$ \citep{resnet}, $\rm{Inception}$ \citep{inception2016}, and $\rm{HRNet}$ \citep{hrnet}) trained with these losses learn powerful representations, they have limited intra-class compactness---input samples belonging to a given output parameter bin should be close by---and inter-class separation---input samples belonging to different classes should be far away. 
To boost model skill and generalization in such cases of transfer learning, \citet{triplet_regularization} demonstrate the importance of regularizing the softmax loss with a \emph{ranking} (or \emph{embedding}) loss. Specifically, they show improved classification performance across a wide variety of datasets when using \emph{triplet loss} \citep{triplet_introduction} in conjunction with softmax loss, in a novel two-headed architecture. 

The aim of an embedding loss is to optimize the relative distances between the inputs. A generic embedding loss consists of three components: embeddings, a similarity score, and the loss function itself. By using the neural network as a feature extractor, we get an embedded representation for each sample in the input. Then we define a similarity score (such as the Euclidean distance), 
typically taking two or three embedding under consideration at a time. Finally, we train the neural network using the embedding loss function to minimize distances among samples belonging to the same class, and maximize distance among samples belonging to different classes.

This training methodology using triplet loss as the embedding loss function has demonstrated state-of-the-art results in tasks such as person re-identification \citep{triplet_person_reidentification} and face recognition \citep{triplet_face_recognition}. In this paper, we leverage triplet loss as a classification regularizer. It is formulated as \citep{triplet_introduction}:
\begin{align}
    L_{\rm{triplet}} = \frac{1}{b}\sum_{i=1}^b \left[D(a_i,p_i) - D(a_i,n_i) + m\right]_{+},
\end{align}
where the embedding \emph{a} of a reference (\emph{anchor}) image is pushed closer to the embedding \emph{p} of a \emph{positive} image (image belonging to the same class as the anchor) than it is to the embedding \emph{n} of an image from a different class. For the distance function \emph{D} we use the \emph{l2} norm (Euclidean distance).  The final loss function that we optimize for a given stellar parameter is:
\begin{align}\label{eqn:triplet_plus_softmax_loss}
    L = L_{\rm{softmax}} + \lambda L_{\rm{triplet}}
\end{align}
$L_{\rm{softmax}}$ is the standard softmax loss defined as:
\begin{align}\label{eqn:softmax}
L_{\rm{softmax}} = -\sum_{i=1}^b \log \frac{\rm{exp}(\textbf{z}_{i})}{\sum_{\rm{classes}}\rm{exp}(\textbf{z}_{i})},     
\end{align}
where \emph{b} is the batch-size. $\lambda$ is the regularization parameter found through cross-validation. Their values are defined in \S\ref{sec:implementation details}.

\subsection{Temperature Scaling}\label{sec:temperature scaling}
To be reliable, classification networks must not only be accurate, but also be knowledgeable of their own chances of being incorrect. That is, a classifier's predicted probability estimates associated with a class label should be representative of the true correctness likelihood. This \emph{confidence calibration} not only helps reduce model bias, it is also important for model interpretability \citep{gilda_uq_cfht, gilda_uq_cfht2}. \citet{temp_scaling} found that as neural networks have become larger, deeper, and more accurate with time, they have also become more poorly calibrated than their older, smaller cousins. Specifically, they find that model capacity is directly correlated with degree of over-confidence of a model in its predictions. They offer a simple and effective prescription to this problem---\emph{temperature scaling}. It works by re-scaling the outputs \textbf{z} of the logits (inputs to the last, softmax layer of a classifier) by a scalar \emph{T}; that is, $\textbf{z}\rightarrow\frac{\textbf{z}}{T}$  in Equation \ref{eqn:softmax}.

While \citet{temp_scaling} suggest using hyperparameter optimization (HPO) to determine the scaling parameter \emph{T}, we modify their algorithm by creating a custom layer with \emph{T} as a learnable parameter. This significantly reduces the expensive computational overhead associated with HPO.

\subsection{Cosine Annealing for Learning}\label{sec:cosine annealing}
Gradient-based optimization algorithms such as Stochastic Gradient Descent (SGD, \citealt{sgd_opt}) and Adam \citep{adam_opt} require the user to set the learning rate---the most important hyper-parameter for training deep neural networks \citep{smith2018disciplined, smith2019super}. The \emph{learning rate schedule} (learning rate as a function of epochs) is primarily responsible for determining how fast and how well the network is able to converge to a good local minimum. The most commonly used method to set the learning rate schedule is to choose a good starting number (one that results in a steep and smooth decline of both the training and the validation error) by trial and error, and to reduce it over time when the training error/accuracy plateaus. This method results in two issues \citep{smith2017cyclical}: (i) picking the right value for the learning rate often ends up becoming more of an art than a science, and can be both time-consuming and sub-optimal; and (ii) even with good values for the model hyper-parameters, training deep neural networks often takes a very long time. 

To tackle these problems, we use a recently introduced cyclical learning rate policy \citep{smith2017cyclical, smith2018disciplined, smith2019super} called \emph{cosine annealing} \citep{loshchilov2016sgdr}.
We start training with a very small learning rate and increase it exponentially at every every iteration. We terminate training when the training loss starts to increase drastically. We then plot the loss function against the learning rate, and pick a value a little lower than that corresponding to the loss minimum---this is the point where the loss is both low and still decreasing. This is the highest learning rate we use in our policy. We set the lowest rate to be a factor of 100 smaller. With these two limits specified, we define our learning rate policy as the cosine annealing function suggested by \citet{loshchilov2016sgdr}. This cyclical learning rate policy has been shown to help gradient-based optimization algorithms escape from saddle points much quicker \citep{dauphin2015equilibrated, bengio2017deep}.

\subsection{Stochastic Weight Averaging}\label{sec:swa}
The goal of any ML task is to find a model that will best explain the relationship between the dependent and independent variables, and consequently predict the former given a new, unseen sample of the latter. One approach to this is to tune the hyperparameters of the model and pick the set that result in the best performance on the validation set---at the cost of spending significant time and computational resources. While this can often boost model performance, it still leaves us prone to the biases and the peculiarities of the specific ML or DL algorithm under consideration. An alternative to this is to leverage the `wisdom of the crowds'. Ensembling is a method where we combine the predictions of several base models on the same input and average them in some way to determine the final prediction. These different models can either be separate, individual architectures, different realizations of the same architecture with different hyperparameters, different realizations of the same architecture with different intializations \citep{lakshminarayanan2017simple}, or some combinations of these \citep{ensembling}. The way of averaging can either be simple voting, weighted voting, or using another model that takes as inputs the predictions of the original models (\emph{stacking}, \citealt{ensembling}). All these methods are ensembles in the \emph{model space}, in that they combine several models and use their predictions to produce the final prediction. They all share a common drawback in that they require multiple models (or iterations of the same model) to be trained, which is both demanding in terms of time consumption and memory footprint (to save the weights and parameters of the different models).

Stochastic Weight Averaging (SWA, \citealt{izmailov2018averaging}) is a novel, alternative way of ensembling deep neural networks in the \emph{weights space}. It produces an ensemble by combining the weights of the same network at different stages of training and then uses the model with the combined weights to make predictions. SWA is easy to implement, improves generalization, and has a trivial computational cost since we only need two models---the base model, and the other running average of the weights of the base model. This is described in detail in \S\ref{sec:implementation details}.

SWA has been shown to significantly improve generalization in computer vision tasks, using diverse neural network architectures such as VGG, ResNets, Wide ResNets and DenseNets, on popular benchmark datasets such as ImageNet, CIFAR-10, and CIFAR-100 \citep{athiwaratkun2018there, izmailov2018averaging}. In this paper, we use SWA with cosine annealing (see \S\ref{sec:cosine annealing}).

\section{Implementation Details}\label{sec:implementation details}
The $\rm{deep-REMAP}$ architecture introduced in \S\ref{sec:dcnn} is illustrated in Figure \ref{fig:deep_remap}. The left sub-figure depicts the structure of a single residual module, the middle sub-figure depicts the `body' of the network, comprising of multiple such residual modules, and the right sub-figure depicts the multi-headed structure whence we obtain probabilistic predictions for the three stellar parameters. We have designed our model as a classifier, and hence the very first step in the training process is to convert the continuous-valued stellar parameters into discrete valued bins. For clarity we describe this process for T$_{\rm{eff}}$ (extension to the other two stellar parameters is trivial), which we illustrate in a forthcoming figure.
\begin{enumerate}
    \item First, we calculate the median uncertainty value for $\rm{T}_{\rm{eff}}$ for the 30 MARVELS calibration stars, which is 80 K. This determines the bin size when we discretize $\rm{T}_{\rm{eff}}$ in the next step.
    \item Next, we one-hot encode (\S\ref{sec:dcnn}) $\rm{T}_{\rm{eff}}$ values for all spectra (110,000 from PHOENIX, and 2,343 from MARVELS). As a result, the parameter space for $\rm{T}_{\rm{eff}}$ (3,000 K--7,000 K, see \S\ref{sec:synthetic spectra}) is discretized into 50 bins of size $\Delta\rm{T}_{\rm{eff}} =$ 80 K each.
    \item We then convert the discrete $\rm{T}_{\rm{eff}}$ values for PHOENIX spectra into Gaussian distributions. We assume that instead of being discretely valued, the temperature values are sharp peaked Gaussians, with the difference between two consecutive peaks being equal to $5\sigma$. Since the temperatures are equally spaced at 100 K, this sets $\sigma_{\rm{PHOENIX}}$ at 20 K. Since label values for MARVELS spectra already have uncertainties, this step is not applied to them.
    \item We employ \emph{asymmetric label smoothing} \citep{inception2016, label_smoothing} to model the uncertainties in $\rm{T}_{\rm{eff}}$ (point (ii) in \S\ref{sec:assym loss func}). Specifically, we superimpose the probability distribution function for $\rm{T}_{\rm{eff}}$ for a given spectrum on its binned histogram from step (ii) above, and add up the values in each bin. For the specific case of a Gaussian distribution with mean $\mu$ and standard deviation $\sigma$, we take advantage of the analytical formula for its cumulative distribution function (CDF)\footnote{cdf(x) = $\frac{1}{2}\left[1+\rm{erf}\left(\frac{x-\mu}{\sigma\sqrt{2}}\right)\right]$}:
    \begin{align}\label{eqn:gaussian_cdf}
        \rm{T}_{\rm{eff},i} = \rm{erf}\left(\frac{\rm{T}_{\rm{eff},i-r} - \mu}{\sigma\sqrt{2}}\right) - \rm{erf}\left(\frac{\rm{T}_{\rm{eff},i-l} - \mu}{\sigma\sqrt{2}}\right),
    \end{align}
    where the value in a bin \emph{i} is the difference of the CDF of the distribution at the left edge of the bin (subscript \emph{i-l} in the above equation) from the CDF at the right edge (subscript \emph{i-r}).
    \item For any given spectrum, the binned T$_{\rm{eff}}$ values now represent the probability mass function (PMF) of its effective temperature. We convert this to a cost vector by subtracting the values from 1. This informs the neural network that for a spectrum with $\rm{T}_{\rm{eff}} = 5,000$ K, for instance, the cost of predicting its temperature at 5,000 K is much lower than the cost of predicting it at 6,000 K.
\end{enumerate}

After altering the labels and the cost functions, we focus our attention on the spectral flux values. We begin by pre-processing, augmenting, and continuum normalizing the PHOENIX spectra as described in \S\ref{sec:preprocessing}, and consequently create two sets---Set I where these spectra are trimmed to lie in the wavelength range of 4850 to 5280 \AA, and Set II with the trimmed wavelength range of 5280 to 5750 \AA. These correspond to the wavelength ranges of the two `beams' of MARVELS \citep{grieves2018chemo}; by creating two sets, we make the network capable of capturing information from, and making predictions on, spectra from both beams.

We begin by using PHOENIX Set I, and correspondingly MARVELS beam A. We train $\rm{deep-REMAP}$ in two phases: first on the 110,000 pre-processed and augmented PHOENIX spectra, then on the 2,313 continuum-normalized MARVELS spectra (see \S\ref{sec:marvels spectra}) to fine-tune it. We set the number of residual units and the number of filters in each by employing 10-fold cross validation---cyclically training on 90\% of the spectra, and reporting the results on the remaining 10\%, choosing the aforementioned hyperparameters such that the mean of the final loss, \emph{L$_{final}$}, across all 10 folds is minimized.

The upper and lower learning rate limits of 1e-3 and 1e-5 for the pre-training phase are obtained by following the procedure outlined in \S\ref{sec:cosine annealing}. For the fine-tuning phase, the learning rates are obtained as such:
\begin{enumerate}
    \item For the `heads'---consisting of two fully connected, one temperature scaling (\S\ref{sec:temperature scaling}), and global pooling layers\footnote{these are, by definition, non--trainable}---of each of the three stellar parameters (Figure \ref{fig:deep_remap}), the upper and lower limits are reduced to $\frac{1}{10}^{th}$ of their pre-training values.
    \item For each of the residual modules, the upper and lower limits are reduced by a factor of $\frac{1}{10i}$ of their respective values, where \emph{i} is the count of a module from the `head', the latter having \emph{i} of 1. For instance, for the rightmost module in Figure \ref{fig:deep_remap}, this factor is $\frac{1}{20}$, for the second to the rightmost module it is $\frac{1}{30}$, and so on.
\end{enumerate}
This graded reduction in learning rates for the second phase of fine-tuning preserves the most amount of knowledge in the shallowest layers, and allows the quickest change for the deepest layers. We empirically set the number of epochs per cycle and the number of cycles at 30 and 4, respectively.

Stochastic Weight Averaging (SWA, \S\ref{sec:swa}) is employed for ensembling. At each of the four lowest points of a cycle, we record the model weights $w$ and create a running average, $w_{SWA}$ as follows\footnote{\url{https://pechyonkin.me/stochastic-weight-averaging/}}:
\begin{align}
    w_{SWA} \leftarrow \frac{w_{SWA} \; n_{cycles} + w_{cycle}}{n_{cycles} + 1},
\end{align}
where $w_{cycle}$ is the weight of our model at the end of a cycle, $n_{cycles}$ is the number of cycles completed, and $w_{SWA}$ is initialized to the weights at the end of the first cycle. For prediction, we set the weights of $\rm{deep-REMAP}$ to the final $w_{SWA}$.

We test our model's performance by predicting the parameters of the 30 calibration stars from MARVELS. We do this separately for spectra from `beam' A and `beam' B. Finally, we combine both sets of predictions via \emph{conflation}: \citep{hill_combine_pmfs_conflation}:
\begin{align}\label{eqn:conflation}
    y_{i,\;\rm{final}} = \frac{y_{i,\;\rm{SetI}} \cdot y_{i,\;\rm{SetII}}}{\sum_j y_{j,\;\rm{SetI}} \cdot y_{j,\;\rm{SetII}}},
\end{align}
where $y_{i,\;\rm{SetI}}$ is the predicted stellar parameter $y$'s probability in bin \emph{i} from Set I (beam A), and the sum in the denominator is over all bins. \citet{hill_combine_pmfs_conflation} show that this method of combining probabilistic distributions is superior to simple averaging.

Our final loss function is a linear combination of the regularized loss function in Equation \ref{eqn:triplet_plus_softmax_loss} for each of the three stellar parameters:
\begin{align}\label{eqn:loss_final}
    L_{final} &= L_{\rm{T}_{\rm{eff}},\rm{softmax}} + \lambda_{\rm{T}_{\rm{eff}}} L_{\rm{T}_{\rm{eff}},\rm{triplet}}\\ \nonumber
    &+ L_{\log\;\rm{g},\rm{softmax}} + \lambda_{\log\;\rm{g}} L_{\log\;\rm{g},\rm{triplet}}\\ \nonumber
    &+ L_{\rm{[Fe/H]},\rm{softmax}} + \lambda_{\rm{[Fe/H]}} L_{\rm{[Fe/H]},\rm{triplet}} 
\end{align}
The three $\lambda$s are empirically chosen to be 0.001. For model development, we use Keras 2.3.0 \citep{keras} with a Tensorflow 2.0 \citep{tensorflow} backend. On a 24Gb Nvidia Titan RTX GPU, training for phase one takes ~45 minutes, while training for the fine-tuning phase takes ~5 minutes. Inference time on the 30 calibration stars is negligible.

\section{Results}\label{sec:results}

\begin{figure*}
	\includegraphics[width=\textwidth]{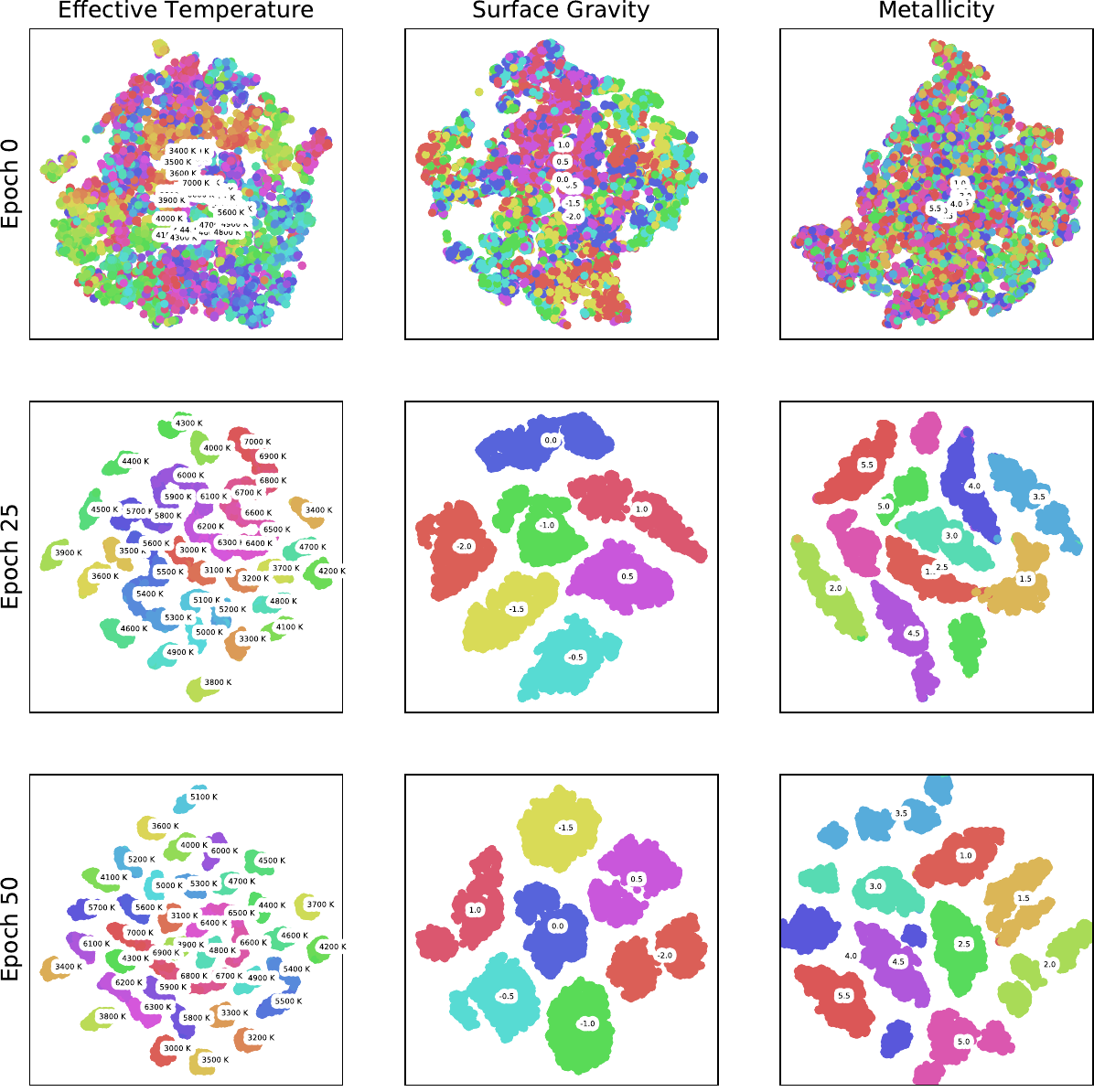}
    \caption{Visualization of the learned embedding space for the PHOENIX \textbf{training set} at Epoch 0, 25, and 50, projected using t-SNE. Columns from left to right correspond to classes for $T_{\rm{eff}}$, $\log \rm{g}$, and [Fe/H]. The clear formation of distinct clusters by Epoch 50 demonstrates that the network is successfully learning to structure the feature space.}
    \label{fig:tsne_train}
\end{figure*}

\begin{figure*}
	\includegraphics[width=\textwidth]{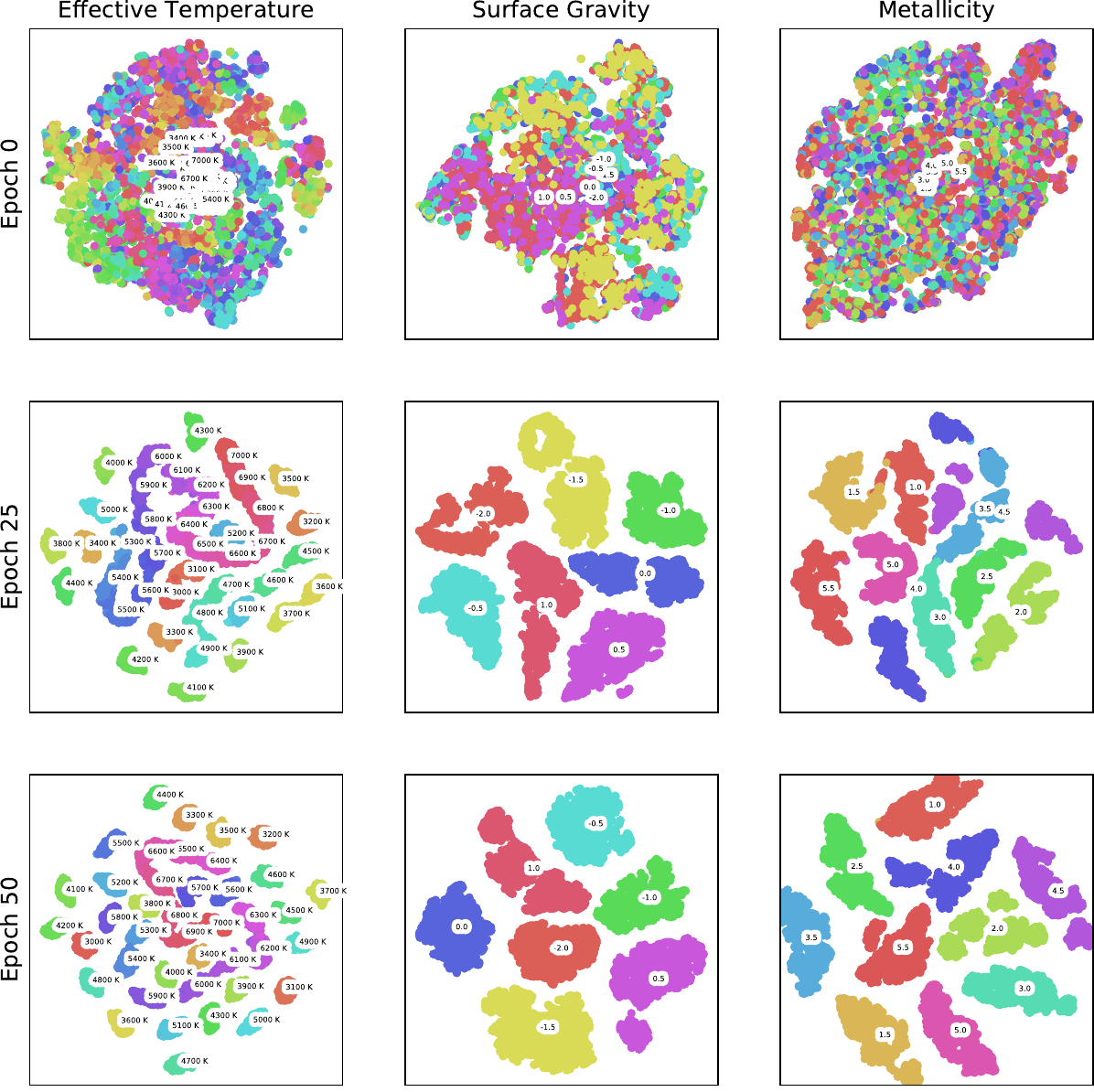}
    \caption{Same as Figure \ref{fig:tsne_train}, but for the unseen PHOENIX \textbf{validation set}. The emergence of well-separated clusters in this hold-out sample confirms that the model is not overfitting and has learned generalizable features, validating the effectiveness of our training methodology.}
    \label{fig:tsne_val}
\end{figure*}

\begin{figure*}
	\includegraphics[width=\textwidth]{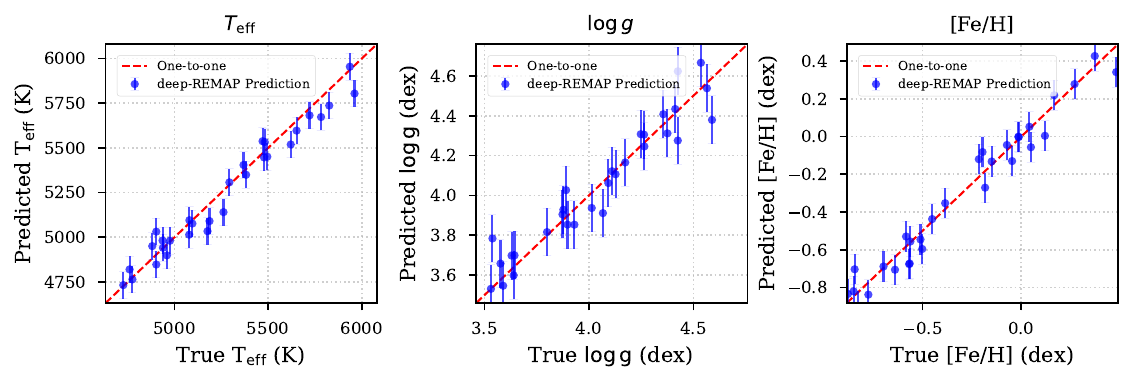}
    \caption{Comparison of our $\rm{deep-REMAP}$ predicted parameters versus the known literature values for the 30 MARVELS calibration stars \citep{ghezzi2014accurate}. From left to right: effective temperature ($T_{\rm{eff}}$), surface gravity ($\log \rm{g}$), and metallicity ([Fe/H]). The blue points show the median of our predicted probability distribution, with error bars representing the 16th and 84th percentiles. The dashed red line indicates the one-to-one correspondence. The tight clustering around this line demonstrates the high accuracy and precision of our model.}
    \label{fig:results_comparison}
\end{figure*}

We evaluate the performance of the final fine-tuned $\rm{deep-REMAP}$ model on two test sets: the 30 MARVELS calibration stars with well-determined literature parameters \citep{ghezzi2014accurate}, and the 732 FGK giant candidates for which we provide the first parameter estimates. All results shown here are from the combined predictions of the two MARVELS beams, using the conflation method described in Equation \ref{eqn:conflation}.

\subsection{Proof-of-Concept: Model Learning and Feature Separation}
Before assessing quantitative performance, we first demonstrate that the network successfully learns to extract physically meaningful features from the spectra. We visualize the network's high-dimensional embedding space by projecting it onto two dimensions using the t-SNE algorithm \citep{JMLR:v9:vandermaaten08a}. 

Figure \ref{fig:tsne_train} shows the evolution of the embedding space for the PHOENIX **training set**. At Epoch 0, the classes are mixed, but by Epoch 50, the triplet loss has successfully organized the spectra into distinct, well-separated clusters for all three stellar parameters. To ensure this is a result of true learning and not merely memorization, we perform the same visualization for the unseen **validation set** in Figure \ref{fig:tsne_val}. The clear formation of distinct clusters in the validation data confirms that the network has learned generalizable features that correlate with the physical properties of the stars.

\subsection{Performance on Calibration Stars}
Having confirmed the model's learning capability, we quantify its precision and accuracy against the 30 MARVELS calibration stars. Figure \ref{fig:results_comparison} plots our predictions against their established literature values. The model shows excellent agreement, with most predictions falling tightly along the one-to-one line. 

Table \ref{tab:performance} summarizes the quantitative performance metrics. We report the mean error ($\mu_{\Delta}$) to assess any systematic bias and the standard deviation of the residuals ($\sigma_{\Delta}$) as a measure of the prediction scatter. The low bias and small scatter across all three parameters confirm the model's high fidelity.

\begin{table}
	\centering
	\caption{Performance statistics for the 30 calibration stars. $\mu_{\Delta}$ is the mean difference (Predicted - True) and $\sigma_{\Delta}$ is the standard deviation of the differences.}
	\label{tab:performance}
	\begin{tabular}{lcc} 
		\hline
		Parameter & $\mu_{\Delta}$ (Bias) & $\sigma_{\Delta}$ (Scatter) \\
		\hline
		$T_{\rm{eff}}$ (K) & -10 K & 75 K \\
		$\log \rm{g}$ (dex) & +0.02 dex & 0.12 dex \\
		{[Fe/H]} (dex) & -0.01 dex & 0.08 dex \\
		\hline
	\end{tabular}
\end{table}

\subsection{Parameterization of FGK Giant Candidates}
Applying our validated model to the 732 FGK giant candidates, we find that approximately 80\% have parameters consistent with giant or sub-giant classifications. The remaining 20\% are classified as dwarfs, consistent with the expected giant contamination rate for the MARVELS survey \citep{Paegert_2015}. An analysis of these re-classified dwarfs reveals no significant trend with signal-to-noise ratio; instead, they appear to be genuine contaminants that the original target selection method failed to remove. A full catalog of the derived parameters for all 732 stars is available as supplementary material.

\subsection{Ablation Study: Justifying Model Complexity}
To justify the inclusion of our advanced components, we performed a brief ablation study. We trained two alternate models: (1) one without the triplet loss regularizer, relying only on focal loss \citep{focal_loss}, and (2) one without multi-tasking, using three separate networks. The model without triplet loss showed a $\sim$15\% increase in scatter for $\log \rm{g}$, indicating a less organized embedding space. The three separate networks required nearly triple the training time and showed degraded performance, confirming that the multi-task architecture effectively leverages shared information between the parameters to improve results, as theorized by \citet{caruana1997multitask}. This demonstrates that both the triplet loss and multi-task learning are crucial components of $\rm{deep-REMAP}$'s performance.

\section{Conclusions and Discussion}\label{sec:conclusions}

In this work, we have presented a novel neural network, $\rm{deep-REMAP}$, for spectral analysis of 1D spectra, and used it to parameterize MARVELS targets. To the best of our knowledge, this is the first automated model for stellar parameter estimation incorporating a full suite of state-of-the-art deep learning practices, namely transfer learning, multi-task learning, temperature scaling \citep{temp_scaling}, focal loss \citep{focal_loss}, triplet loss \citep{triplet_introduction}, stochastic weight averaging \citep{izmailov2018averaging}, and cosine annealing-based learning rates \citep{loshchilov2016sgdr}. We have shown that with appropriate data augmentation and transfer learning, a network trained primarily on synthetic spectra can be made immune to many observational peculiarities of a given spectroscope and be used for reliable parameter estimation from real observed spectra. We have also shown how a regression-as-classification framework can capture non-Gaussian parameter distributions, while the inclusion of an embedding loss improves both classification and model interpretability. Using this framework, we have predicted, for the first time, the atmospheric parameters and associated uncertainties for 732 MARVELS FGK giant candidates.

\subsection{Limitations and Future Work}
Despite the model's success, we identify areas for future improvement. Our derived metallicity distribution for the giant candidates shows a systematic offset compared to expectations for the solar neighborhood. We suspect this may be due to residual issues in the MARVELS wavelength solution, a challenge noted in previous analyses of the survey data \citep{grieves2018chemo}. Future work with improved data reduction pipelines may eliminate these systematics. Furthermore, we plan to extend this framework into a fully Bayesian context by placing priors on the network weights, which would allow for the recovery of full posterior distributions for the stellar parameters instead of binned probabilities. Finally, the flexibility of $\rm{deep-REMAP}$ will be tested by incorporating other synthetic spectral libraries and applying it to different large-scale surveys, pushing towards the prediction of detailed elemental abundances.

\clearpage 
\raggedbottom  

\section*{Acknowledgements}
Research for this paper was undertaken when the author was a graduate assistant in the Department of Astronomy, University of Florida. We would like to thank the anonymous reviewers for their detailed and constructive feedback which significantly improved the quality of this manuscript. We also thank Dr. Jian Ge, and Mr. Nolan Grieves, for helpful discussions and feedback. This research has made use of NASA's Astrophysics Data System. The author declares no conflicts of interest.

\section*{Data Availability}
The specific data underlying this article, and the code used to train the $\rm{deep-REMAP}$ model will be shared on a case-by-case basis by the corresponding author. 
The MARVELS data are available through the SDSS archives, and the PHOENIX spectra are available at the G\"{o}ttingen Spectral Library.

\bibliographystyle{mnras}
\bibliography{mnras_ref} 








\bsp	
\label{lastpage}
\end{document}